\documentclass[twocolumn]{aastex62}

\newcommand{\Kepler}{{\it Kepler}}

\newcommand{\Gaia}{{\it Gaia}}
\newcommand{\be}{\begin{equation}}
\newcommand{\ee}{\end{equation}}

\newcommand{\metallicity}{[Fe/H]}
\newcommand{\msun}{M$_\odot$}
\newcommand{\rsun}{R$_\odot$}

\newcommand{\kms}{\ensuremath{\rm km\,s^{-1}}}

\newcommand{\loggspc}{3.94}
\newcommand{\loggespc}{0.44}

\newcommand{\mh}{-0.10}
\newcommand{\mhe}{0.08}

\newcommand{\mstar}{1.09}
\newcommand{\mstare}{0.09}

\newcommand{\rstar}{1.24}
\newcommand{\rstare}{0.10}

\newcommand{\rearth}{R$_\oplus$}

\newcommand{\ldone}{0.316}
\newcommand{\uldone}{0.097}
\newcommand{\ldtwo}{0.394}
\newcommand{\uldtwo}{ 0.314}
\newcommand{\rprstb}{0.0742}
\newcommand{\urprstb}{0.0039}
\newcommand{\arstb}{7.94}
\newcommand{\uarstb}{0.55}
\newcommand{\inclb}{88.44}
\newcommand{\uinclb}{^{+1.09}_{-1.91}}
\newcommand{\impb}{0.21}
\newcommand{\uimpb}{^{+0.23}_{-0.15}}
\newcommand{\rplb}{9.916}
\newcommand{\urplb}{0.985}
\newcommand{\rplbl}{0.884}
\newcommand{\urplbl}{0.087}
\newcommand{\perplb}{3.38628} 
\newcommand{\uperplb}{2\times10^{-5}}
\newcommand{\ttransitb}{2457823.2233}
\newcommand{\uttransitb}{0.0006}

\newcommand{\tdurb}{3.09}
\newcommand{\utdurb}{^{+0.26}_{-0.04}}

\newcommand{\thisstar}{EPIC~246865365}
\newcommand{\thisplanet}{EPIC~246865365b}
\usepackage{xcolor}

\definecolor{my_color}{HTML}{3a18b1}

\definecolor{new_color}{HTML}{CF0000}

\accepted{June 4th, 2019}
\submitjournal{\textit{The Astronomical Journal}}

\shorttitle{A hot Saturn near NGC 1817}
\shortauthors{Rampalli et al.}

\begin{document}

\title{A Hot Saturn Near (but unassociated with) the Open Cluster NGC 1817}

\correspondingauthor{Rayna Rampalli}
\email{rmr2196@columbia.edu}

\author[0000-0001-7337-5936]{Rayna Rampalli}
\affil{Department of Astronomy, Columbia University, 550 West 120th Street, New York, NY 10027, USA}

\author[0000-0001-7246-5438]{Andrew Vanderburg}
\altaffiliation{NASA Sagan Fellow}
\affiliation{Department of Astronomy, The University of Texas at Austin, Austin, TX 78712, USA}

\author[0000-0001-6637-5401]{Allyson Bieryla}
\affiliation{Center for Astrophysics $\vert$ Harvard $\&$ Smithsonian, 60 Garden St, Cambridge, MA, 02138, USA}

\author[0000-0001-9911-7388]{David W. Latham}
\affiliation{Center for Astrophysics $\vert$ Harvard $\&$ Smithsonian, 60 Garden St, Cambridge, MA, 02138, USA}

\author[0000-0002-8964-8377]{Samuel N. Quinn}
\affiliation{Center for Astrophysics $\vert$ Harvard $\&$ Smithsonian, 60 Garden St, Cambridge, MA, 02138, USA}

\author[0000-0002-1917-9157]{Christoph Baranec}
\affiliation{Institute for Astronomy, University of Hawai‘i at M\={a}noa, Hilo, HI 96720-2700, USA}

\author{Perry Berlind}
\affiliation{Center for Astrophysics $\vert$ Harvard $\&$ Smithsonian, 60 Garden St, Cambridge, MA, 02138, USA}

\author[0000-0002-7714-6310]{Michael L. Calkins}
\affiliation{Center for Astrophysics $\vert$ Harvard $\&$ Smithsonian, 60 Garden St, Cambridge, MA, 02138, USA}

\author[0000-0001-9662-3496]{William D. Cochran}
\affiliation{Department of Astronomy, The University of Texas at Austin, Austin, TX 78712, USA}

\author[0000-0001-5060-8733]{Dmitry A. Duev}
\affiliation{Department of Astronomy, California Institute of Technology, 1200 E. California Blvd., Pasadena, CA 91101, USA}

\author[0000-0002-7714-6310]{Michael Endl}
\affiliation{Department of Astronomy, The University of Texas at Austin, Austin, TX 78712, USA}

\author[0000-0002-9789-5474]{Gilbert A. Esquerdo}
\affiliation{Center for Astrophysics $\vert$ Harvard $\&$ Smithsonian, 60 Garden St, Cambridge, MA, 02138, USA}

\author[0000-0003-0054-2953]{Rebecca Jensen-Clem}
\affiliation{Astronomy Department, University of California Berkeley, Berkeley, CA 94720-3411, USA}

\author[0000-0001-9380-6457]{Nicholas M. Law}
\affiliation{Department of Physics and Astronomy, University of North Carolina at Chapel Hill, Chapel Hill, NC 27599-3255, USA}

\author[0000-0002-7216-2135]{Andrew W. Mayo}
\affiliation{Astronomy Department, University of California Berkeley, Berkeley, CA 94720-3411, USA}

\author[0000-0002-0387-370X]{Reed Riddle}
\affiliation{Department of Astronomy, California Institute of Technology, 1200 E. California Blvd., Pasadena, CA 91101, USA}

\author[0000-0002-5082-6332]{Ma{\"i}ssa Salama}
\affiliation{Institute for Astronomy, University of Hawai‘i at M\={a}noa, Hilo, HI 96720-2700, USA}



\begin{abstract}
We report on the discovery of a hot Saturn-sized planet (\rplb $\pm$\urplb\ \rearth) around a late F star, EPIC 246865365, observed in Campaign 13 of the \textit{K2} mission. We began studying this planet candidate because prior to the release of \Gaia\ DR2, the host star was thought to have been a member ($\geq90\%$ membership probability) of the $\approx1$~Gyr open cluster NGC 1817 based on its kinematics and photometric distance. We identify the host star (among three stars within the \textit{K2} photometric aperture) using seeing-limited photometry and rule out false positive scenarios using adaptive optics imaging and radial velocity observations. We statistically validate \thisplanet\ by calculating a false positive probability rate of $0.01\%$. However, we also show using new kinematic measurements provided by \Gaia\ DR2 and our measured radial velocity of the system that EPIC 246865365 is unassociated with the cluster NGC 1817. Therefore, the long-running search for a giant transiting planet in an open cluster remains fruitless. Finally, we note that our use of seeing-limited photometry is a good demonstration of similar techniques that are already being used to follow up \textit{TESS} planet candidates, especially in crowded regions.

\end{abstract}


\section{Introduction}

Open clusters have long served as benchmarks for studying stellar, dynamical, and most recently, planet, evolution due to their homogenous nature. These populations provide a sample of stars of approximately the same age, distance, and metallicity over a large range in mass. As a result, when studying how exoplanets depend on their host stellar properties, clusters are ideal environments to target.

Additionally, since open clusters are often made up of young stellar populations, studying planets in open clusters is imperative for characterizing planetary evolution since the most formative time for planets is thought to be in the first gigayear (e.g. \citealt{2010Mann} , \citealt{2012Lopez}). For example, on this timescale, the increased stellar rotation and magnetic activity lead to an excess of X-ray and ultraviolet emission that could erode atmospheres of close-in planets \citep{Lammer}. Comparing planets found in the younger stellar populations to those found around older field stars could illuminate just exactly how planetary systems evolve.

There have been a number of planet surveys targeting clusters, and a handful of planets have been discovered. Through radial velocity (RV) surveys, several hot Jupiters were discovered in the Hyades and Praesepe, both around 650~Myrs in age (\citealt{2007Sato}; \citealt{2012Quinn}; \citealt{2014Quinn}). The \textit{\textit{K2}} mission \citep{howell} has observed a number of clusters (including the Hyades, Praesepe, and others) and has yielded close-in, Neptune-sized and smaller planet detections (e.g. \citealt{Crossfield}; \citealt{2016Mann}; \citealt{2017Pepper}; \citealt{2018Vanderburg}; \citealt{ruprecht}). However, there have not been any transiting giant planets found around clusters, most likely because hot Jupiters are intrinsically rare \citep{vansadersgaudi}. The largest transiting planets in clusters found to date are Neptune-sized planets orbiting stars in NGC 6811 \citep{meibom}.  

NGC 1817 is an open cluster observed by \textit{\textit{K2}} in its thirteenth campaign. Its age is between 0.8 to 1.2~Gyrs, similar to that of both the Hyades and Praesepe with a subsolar metallicity of [Fe/H] $=$ $-$0.40~dex \citep{2013Donati}. Located in the constellation of Taurus, 1.7~kpc from Earth, it is 9.6~kpc from the galactic center with a proper motion of 0.485 RA, $-$0.89 Dec mas yr\textsuperscript{-1} and an RV of 65.3 $\pm$ 0.1 km s\textsuperscript{-1} (\citealt{2018GaiaCluster}, \citealt{2004balaguer}).

In this paper, we initially identified a giant planet candidate transiting a star which had been considered a possible member of NGC 1817, potentially making it the first discovery of a giant transiting planet in an open cluster. Via follow-up observations and modeling, we were able to validate the candidate as a genuine transiting planet, but our analysis showed that the host star is not a cluster member. In Section 2, we describe our follow-up observations. In Section 3, we describe our analysis used to determine system parameters, validate the planet and disqualify the star as a cluster member. In Section 4, we discuss the implications of our findings and methods for future planet searches in clusters and conclude.

\begin{figure*}[ht!] 
  \centering
  \includegraphics[width=6.5in]{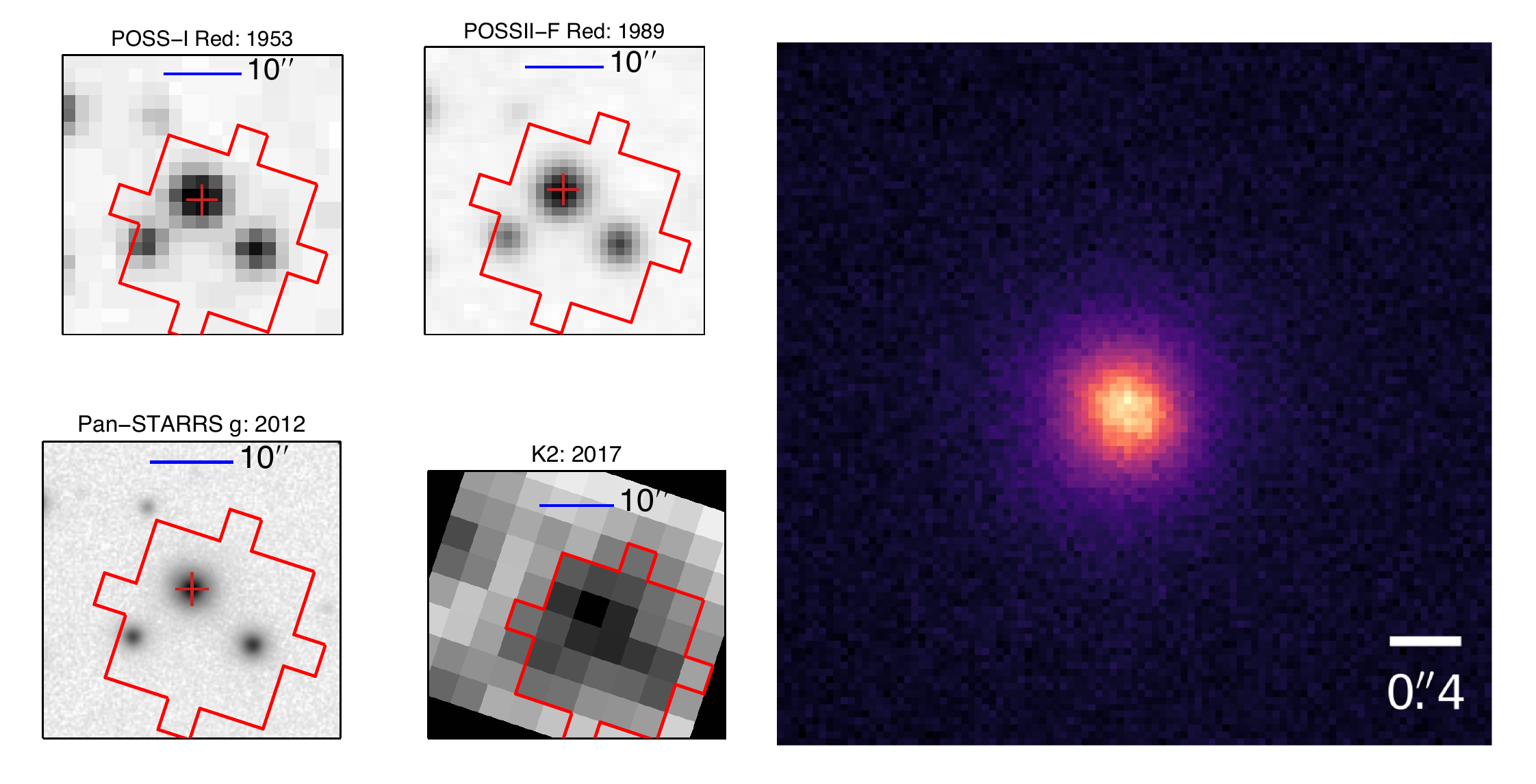}
   \caption{\textit{Top Left}: Image from the first Palomar Sky Survey (POSS-I) with red outline indicating the shape of the aperture chosen for reduction. \textit{Top Middle}: Image from the second Palomar Sky Survey (POSSII). \textit{Bottom Left}: Image from Panoramic Survey Telescope and Rapid Response System (Pan-STARRS). \textit{Bottom Middle}: Image from \textit{K2} Field of View rotated to match orientation of previous 3 images. \textit{Right}: High-resolution image of target obtained with Robo-AO.}
   \label{fov}
\end{figure*}

\section{Observations}\label{observations}

\subsection{\textit{K2} Light Curve and Archival Imaging}\label{lightcurve}
 
EPIC 246865365 was observed from 2017 March 08 to 2017 May 27 by the $\Kepler$ space telescope in the thirteenth campaign of its \textit{\textit{K2}} mission. We downloaded the subsequent calibrated target pixel files (TPFs) and extracted the light curve from the TPF using an aperture of about 20\arcsec\ in diameter. The light curve was processed following \cite{vj14} to produce a photometric light curve free of systematic effects due to the instrument's unstable pointing. We then performed a transit-search using a Box Least Squares Periodogram search (\citealt{kovacs}, \citealt{2016vanderburg}), and the light curve was visually inspected to ascertain with high confidence that the detected events were consistent with real planetary transits. We note that there was no obvious stellar rotation signal in the light curve. We then re-derived the \textit{K2} systematics correction by simultaneously fitting for the transits, \textit{K2} roll systematics, and long-term trends in the dataset, following \citet{2016vanderburg}. We use the light curve with this better-optimized systematics correction in our analysis throughout the rest of the paper. 

\Kepler\ has large (4\arcsec) pixels and a 6\arcsec\ undersampled point spread function, so we downloaded higher-resolution archival images of the region of sky around \thisstar\ from the first and second Palomar Observatory Sky Surveys and the PanSTARRS survey. These images reveal that there are three stars within the photometric aperture we used to extract the \textit{K2} light curve (EPIC 246865365 and two other, slightly fainter stars), any one of which could plausibly be the source of the transit signals (see Figure \ref{fov}).
We later used seeing-limited follow-up photometry to determine the host (described in Section \ref{seeingltd}).

\subsection{Seeing-Limited Follow-up Photometry} \label{seeingltd}
In order to determine which star hosted the signal, we performed seeing-limited follow-up photometry. Seeing-limited photometry can prove useful when needing to rule out nearby eclipsing-binaries as sources of the transit since many ground-based telescopes can achieve higher spatial resolution than focus-limited telescopes like \Kepler. We observed the predicted transit that occurred on UT 2017 November 16 using the 1.2m telescope at the Fred Lawrence Whipple Observatory (FLWO) on Mount Hopkins, Arizona. Images were taken using KeplerCam, a wide-field CCD camera with a 23\farcm1 square FOV and resolution of 0\farcs336 per pixel. The target was observed using the Sloan \textit{i}-band filter with an exposure time of 180 seconds. 

Observing conditions were nominal with a measured seeing of 1\farcs98 (full width at half maximum). Observations of the target began 20 minutes before predicted ingress and ended 2 hours after the predicted egress with an average airmass of 1.18. As a result, observations during the entire 3.2 hour predicted transit duration were obtained. Since KeplerCam is a single-chip CCD read out by 4 amps, the raw images are saved in four sections that we later stitched together with an IDL script. Standard IDL routines were also used to calibrate the images \citep{idlred}. 

Differential aperture photometry was performed using AstroImageJ \citep[AIJ,][]{AIJ}. The AIJ apertures are made up of three concentric circular rings (with sizes chosen based on the measured FWHM seeing for the target) defining three separate regions: an inner circular aperture for measuring the star's flux (with radius $r < $6\arcsec), an annulus acting as a buffer between the inner and outer apertures (6\arcsec$<r<$18\arcsec), and an outer annulus measuring the sky background (18\arcsec$<r<$25\arcsec). In order to determine the source of the transit signal, relative fluxes were measured for the target star, any neighboring stars suspected to contaminate the signal from the target, and comparison stars for the target star and potential contaminant stars. $\text{BJD}_{\text{TDB}}$ timestamps were used in order to remain consistent with data from \Kepler. 

We detected a transit signal consistent in transit depth and timing with the \textit{K2} light curve from the target star while the neighboring stars remained flat throughout the night, confirming the target as the host.

\subsection{Spectroscopy}\label{spectroscopy}

\subsubsection{Tull Spectrograph at McDonald Observatory}
We observed \thisstar\ using the high-resolution cross-dispersed Echelle spectrometer on the Harlan J. Smith 2.7m telescope at McDonald Observatory \citep{1995Tull}. This spectrometer is fed by a 1\farcs2$\times$8\farcs2 slit and has wavelength coverage from 375~nm to 1020~nm. These spectra were obtained on January 28, 29, and February 25, 2018 with a resolving power of $\Delta\lambda/\lambda \approx$ 60,000. Exposure times of 3600 seconds yielded signal-to-noise ratios of about 25 per resolution element at 565~nm.

For each observation, three successive short exposures were taken to remove the energetic particle collisions on the detector. In order to obtain an accurate flux-weighted barycentric correction, an exposure meter was used. The raw data were then processed using IRAF routines for bias subtraction, order extraction, and flat fielding. For each spectral order, apertures were traced, and an extraction algorithm was applied to retrieve the stellar flux as a function of wavelength. 

To calibrate wavelength and remove spectrograph drifting, we obtained bracketing exposures of a Th-Ar hollow cathode lamp, which enabled calculation of absolute RVs from the spectra. We calculate an absolute RV of $\approx 9$ km s\textsuperscript{-1} for this star and calculate stellar parameters from the spectra using \textit{Kea} \citep{2016Endl}, as described in Section \ref{stellarparameters}.

\subsubsection{TRES Spectrograph at Whipple Observatory}
We also observed EPIC 246865365 on January 22 and February 6, 2018 using the Tillinghast Reflector Echelle Spectrograph (TRES) on the 1.5m telescope at FLWO. The spectra were acquired at a spectral resolving power of $\lambda/\Delta\lambda =$ 44,000 in a series of three exposures, each with an exposure time of 1200 seconds. Between these exposures, bracketing exposures of a Th-Ar hollow cathode lamp were taken. However, due to the target's $V$-magnitude of 15, the obtained spectra only yielded signal-to-noise ratios of 12 per resolution element at 518~nm even in good observing conditions. These spectra were too weak to use to derive accurate stellar parameters for the system or to derive RVs with our standard procedure, which uses the strongest observed spectrum of a star as the template for cross-correlation. Instead, we use a similar star from Praesepe as the template to derive RVs for the system; this process is described further in Section \ref{rvanalysis}.

\subsection{Adaptive Optics Imaging}\label{imaging}
We obtained images of EPIC 246865365 (as seen in Figure \ref{fov}) using Robo-AO at the Kitt Peak National Observatory 2.1m telescope (\citealt{Baranec2014}, \citealt{JensenClem2018}). Robo-AO is a robotic laser guide-star adaptive optics system that can be used to determine if there are other potential sources in the \Kepler\ photometric apertures affecting the transit signal (e.g. \citealt{Law2014}). 

Observations were acquired using a long-pass filter that cuts on at $\lambda = 600$~nm on UT 3 March 2018 as a series of frame-transfer exposures at a rate of 8.6 frames per second, for a total time of 120 s. Effective seeing at the time of observation was measured to be approximately 1.6\arcsec\ in a 10 s exposure captured during setup of the adaptive optics system. Because of the poor seeing and faintness of the target, we relied on the faint-star shift-and-add pipeline, described by \citet{JensenClem2018}, to combine the short exposures to maximize the S/N of the final processed image. The final image width of EPIC 246865365 was measured to be 0.43\arcsec, and, from visual inspection, it appears there are no neighboring stars $\lesssim$\,2 magnitudes fainter than \thisstar\ close enough to contaminate its signal.

\section{Analysis}
\label{analysis}

\subsection{Transit Light Curve}\label{transitanalysis}

Due to the three stars in the photometric aperture (Figure \ref{fov}), before measuring transit parameters, we applied a dilution correction to the \textit{K2} light curve. Assuming the flux contained within the aperture is solely from those three stars, we obtained the \Gaia\ $G$-magnitudes (comparable to the \Kepler\ band) for the three stars within the aperture and converted them to fluxes. After confirming the target was indeed the host of the transit signal (see Section \ref{seeingltd}), we determined the fraction of the flux within the aperture coming from the target star and applied the corresponding dilution as shown below: 
\begin{equation}
    F_{new} = \frac{F-(1-d)}{d},
\end{equation}
where $F_{new}$ is the corrected flux, $F$ is the original flux, and $d$ is the fraction of the aperture's flux coming from the target star. For the target, we found $d$ = 0.731. 

We also calculated the dilution using \textit{Kepler's} point spread function (PSF) to model the flux fraction in the aperture and find a $d$ = 0.751. This results in $\approx 3\%$ change in transit depth compared to our initial calculation. However, we find an uncertainty of $\approx 10\%$ for the transit depth in Section \ref{transitanalysis} indicating that this difference in dilution does not significantly change our results.

We measured transit parameters for the system by fitting the \textit{K2} light curve (flattened by dividing away the best-fit low-frequency variability from the simultaneous transit/systematics/low-frequency fit) and KeplerCam light curve with a Markov Chain Monte Carlo (MCMC) algorithm, emcee, \citep{emcee} using model light curves produced by the batman package \citep{batman}. This model follows the \citet{mandelagol} algorithm and was oversampled and binned to match \Kepler's long exposure times as done by \cite{binning}. We also oversampled and binned the model used for the KeplerCam data to account for the 180 second exposure time.

Our model parameters include the planet to star radius ratio $\left(R_{p}/R_{*}\right)$, epoch of the first \textit{K2} transit mid-point $\left(t_{0}\right)$, orbital period, scaled semi-major axis $\left(a/R_{*}\right)$, and orbital inclination. We also fit for normalization coefficients ($C_{1}$,$C_{2}$) for the KeplerCam transit so that both the \textit{K2} and KeplerCam data could be used together for measuring transit parameters. We introduce $C_{1}$ and $C_{2}$ in the following manner to fit for the offset: 
\begin{equation}
    \Theta_{kepcam} = \theta_{kepcam}\left(C_{1}+C_{2}\left(\frac{t-t_{0}}{t_{tot}}\right)\right), 
\end{equation}
where $\Theta_{kepcam}$ is the full forward modeled light curve, $\theta_{kepcam}$ is the light curve model generated by BATMAN, $C_{1}$ and $C_{2}$ are the flux offset and slope, $t$ is time of each KeplerCam flux measurement, $t_{0}$ is the first KeplerCam time stamp, and $t_{tot}$ is the total duration of KeplerCam observations.

Assuming a quadratic limb-darkening law, we impose Gaussian priors on the two limb-darkening coefficients, $u_{1}$ and $u_{2}$ centered on 0.324 and 0.299 following \cite{claretbloemen}. We run the sampling with a fixed eccentricity of 0 and a fixed longitude of periastron of 90 degrees.

\begin{figure*}[ht!] 
  \centering
  \includegraphics[width=7 in]{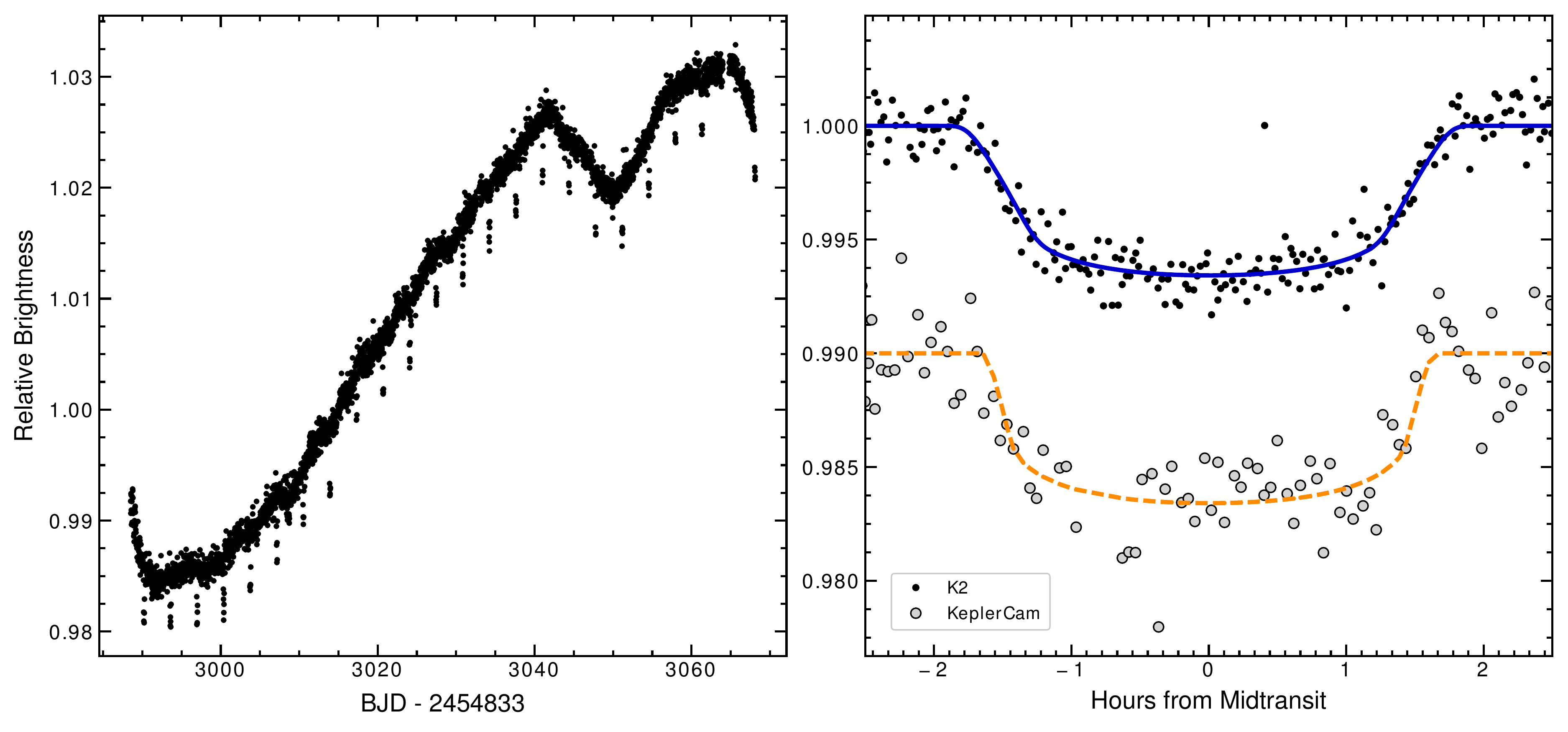}
   \caption{\textit{K2} light curve of \thisstar. 
   \textit{Left}: Full \textit{K2} light curve from \citep{vj14} with systematics removed. \textit{Right}: Phase-folded \textit{K2} light curve (black points) and best-fit transit model (solid blue line). KeplerCam light curve (grey points) and best-fit transit model (dashed orange line) with an imposed relative offset of 0.001 for clarity.}
   \label{lc}
\end{figure*}

We sampled the parameter space with 150 walkers for 50000 steps and discarded the first 5000 as ``burn-in". The transit light curves and best-fit model are shown in Figure \ref{lc}, and transit parameters and uncertainties are listed in Table \ref{bigtable}; for each parameter, we report the median value with errors as the 16th and 84th percentiles corresponding to 1$\sigma$ errors for a Gaussian distribution. 

\subsection{Stellar Parameters}\label{stellarparameters}

We were able to measure spectroscopic parameters of EPIC 246865365 by summing the three McDonald spectra and analyzing the coadded spectrum with the \textit{Kea} software package. \textit{Kea} compares the input spectrum with a large grid of synthetic model stellar spectra to determine spectroscopic properties including stellar effective temperatures, surface gravities, and metallicities \citep{2016Endl}. From the \textit{Kea} analysis, we find a $T_{\text{eff}}$ of 6100 $\pm$ 263~K, metallicity of [Fe/H] = $-$0.10 $\pm$ 0.08~dex, surface gravity of log $g_{\text{cgs,Kea}}$ = 3.94 $\pm$ 0.44~dex, and a projected rotational velocity with an upper limit of $v\sin{i}$ $\leq$ 29 $\pm$ 5 km s\textsuperscript{-1}.

Assuming a circular orbit and using the \textit{K2} transit photometry, we estimate a stellar density with the measured $a/R_{*}$, period, and Kepler's third law following \cite{seagerornelas}. We assumed a zero eccentricity for the system since a majority of short period planets, including \thisplanet\ , have circularization timescales much less than the age of their host stars.\footnote{We confirmed that \thisplanet's circularization timescale is much shorter than the age of the star following \cite{goldreich} and \cite{Mills} Using the 50th percentile of the calculated mass limit distribution (described later in Section \ref{rvanalysis}), we estimate a conservative upper limit on the circularization timescale of 315~Myrs. More reasonable estimates for the planetary mass, assuming it is similar to Saturn, give circularization timescales an order of magnitude shorter, much less than the age of the host star.}

With the stellar density and measured spectral parameters including metallicity, effective temperature, and surface gravity, we then determine the mass, radius, and expected parallax of the star following \cite{determinemasses}. This was done using MCMC with an affine invariant ensemble sampler to explore parameter space in stellar mass, metallicity, and age based on the Yonsei-Yale isochrones \citep{yi}, while imposing priors on the stellar density, metallicity, temperature, and surface gravity from our previous analysis.

We find that EPIC 246865365 has a mass of 1.09 $\pm$ 0.09 \msun\ and a radius of 1.24 $\pm$ 0.10 \rsun, indicating a late F-type dwarf star. Based on the models and V-band brightness of \thisstar, we calculate an expected parallax of $0.66 \pm 0.09$ mas. Our predicted parallax is within 2.5 sigma of the \Gaia\ DR2 value ($-$1.25 $\pm$ 0.72 mas), a nondetection due to the star's faintness and distance.

\subsection{Radial Velocity Analysis}
\label{rvanalysis}
We measured the absolute RV of \thisstar\ from the McDonald spectra using \textit{Kea}. Using observations of RV standard stars taken on the same night, we place the RVs on the IAU absolute velocity scale. 

\begin{figure*}[ht!] 
  \centering
  \includegraphics[width=7.5 in]{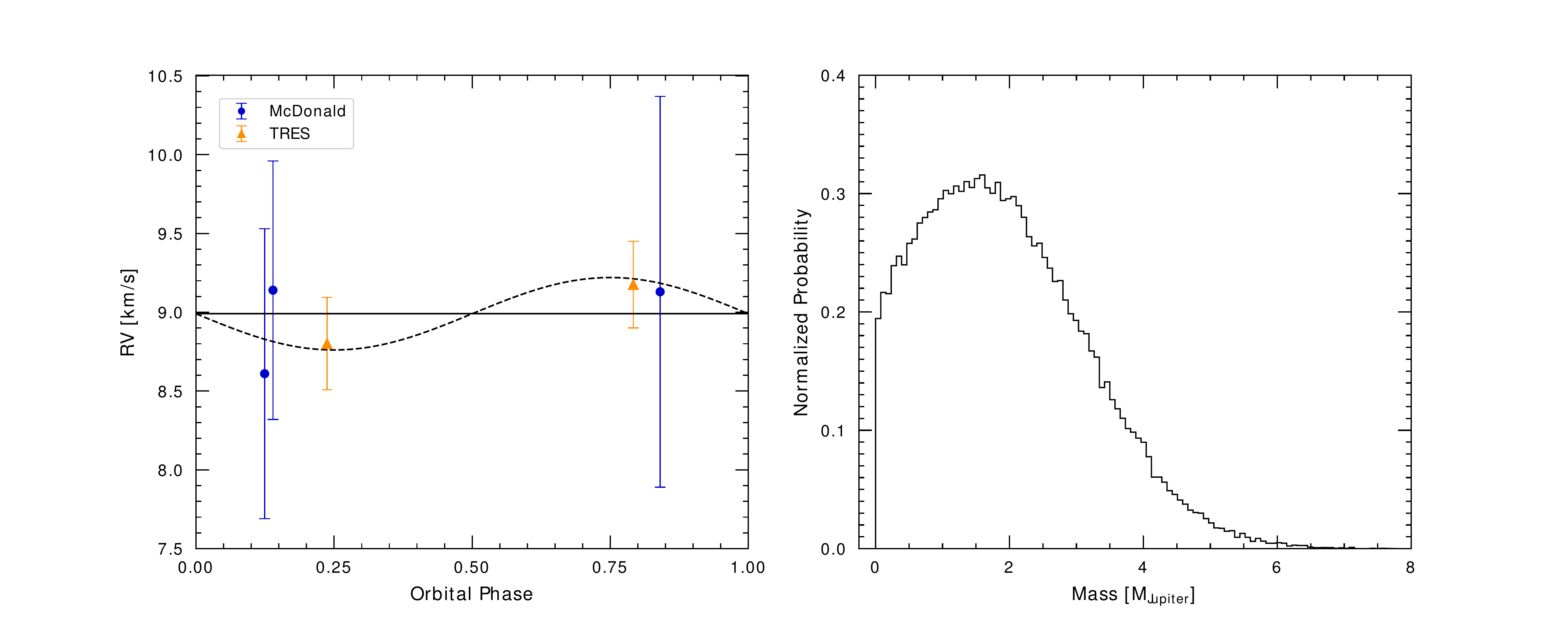}
   \caption{
   \textit{Left}: Absolute RVs as calculated from from McDonald spectra (blue circles) and TRES spectra (orange triangles) as a function of orbital phase.  \textit{Right}: Normalized posterior probability distribution of planet mass limit. Our analysis shows that the mass must be less than 8 Jupiter masses, far below the mass required to be a stellar companion.}
   \label{rv}
\end{figure*}

Because the TRES observations were too weak for us to derive reliable radial velocities with our standard procedures (cross-correlating against a suite of model spectra), we performed a custom analysis. We cross correlated the two TRES spectra against a high SNR TRES spectrum of the Praesepe star EPIC 211926132 ($T_{\text{eff}}$ =6250~K; $v\sin{i}$ = 10 km s\textsuperscript{-1}) and estimated the relative RV of EPIC 211926132 using multiple echelle orders. We measured an $\approx$350 m s\textsuperscript{-1} shift between the two spectra, with uncertainties of 300 m\,s\textsuperscript{-1} velocity. We placed the two TRES velocities on the IAU scale using historical observations of RV standard stars. There were no significant variations when combined with the absolute RVs derived from the McDonald spectra as seen in Figure \ref{rv}. 

We calculated the mass limit of the planet candidate using the RVs from both McDonald and TRES, which we show in Figure \ref{rv}. This was done by phase-folding the RV observations using the measured period from the transit-fitting (Section \ref{transitanalysis}) and fitting a sine curve with emcee to determine the semi-amplitude of the system. Following \cite{rvcalc}, we determined the planet mass limit from the semi-amplitude. Using the calculated stellar mass (described in Section \ref{stellarparameters}) and the distribution of semi-amplitudes calculated by the emcee fit, we calculate a posterior probability distribution for the planet's mass.

Our analysis shows that if the transiting object orbits \thisstar\, its mass is much less than that of a stellar companion. While the use of another star as the template can introduce small systematic errors to the velocities, we can confidently exclude the presence of an eclipsing binary orbiting the host star.

We note that despite obtaining weaker spectra from TRES compared to the spectra from McDonald, we derive absolute RVs with smaller uncertainties. This can be attributed to the better instrumental stability and fiber-fed setup of TRES compared to the slit-fed Tull spectrograph at McDonald. 

\subsection{(Non-)Membership in NGC 1817}
NGC 1817 is an open cluster similar in age to the Hyades located ($\approx$ 1 Gyr) in the constellation of Taurus. Prior to \Gaia\ Data Release 2 (DR2), its cluster members were reported to have a proper motion of (3.56, $-$6.7) mas yr\textsuperscript{-1} in (RA, DEC), and a RV of 65.3 $\pm$ 0.1 km s \textsuperscript{-1} (\citealt{2009Wu}, \citealt{2004balaguer}). Recently, the cluster's proper motion was updated using data from \Gaia\ DR2, giving (0.485, $-$0.89) mas yr\textsuperscript{-1} in (RA,DEC) \citep{2018GaiaCluster}. 

EPIC 246865365 was long considered to be a possible member of the cluster given its similar proper motion of (3.3 $\pm$ 4, $-$1.2 $\pm$ 4) mas yr\textsuperscript{-1} in (RA, DEC) \citep{2010KroneMartins}. In fact, \citet{2010KroneMartins} calculated the star to have a 92.1\% membership probability. 

\begin{figure*}[ht!] 
  \centering
  \includegraphics[width=7 in]{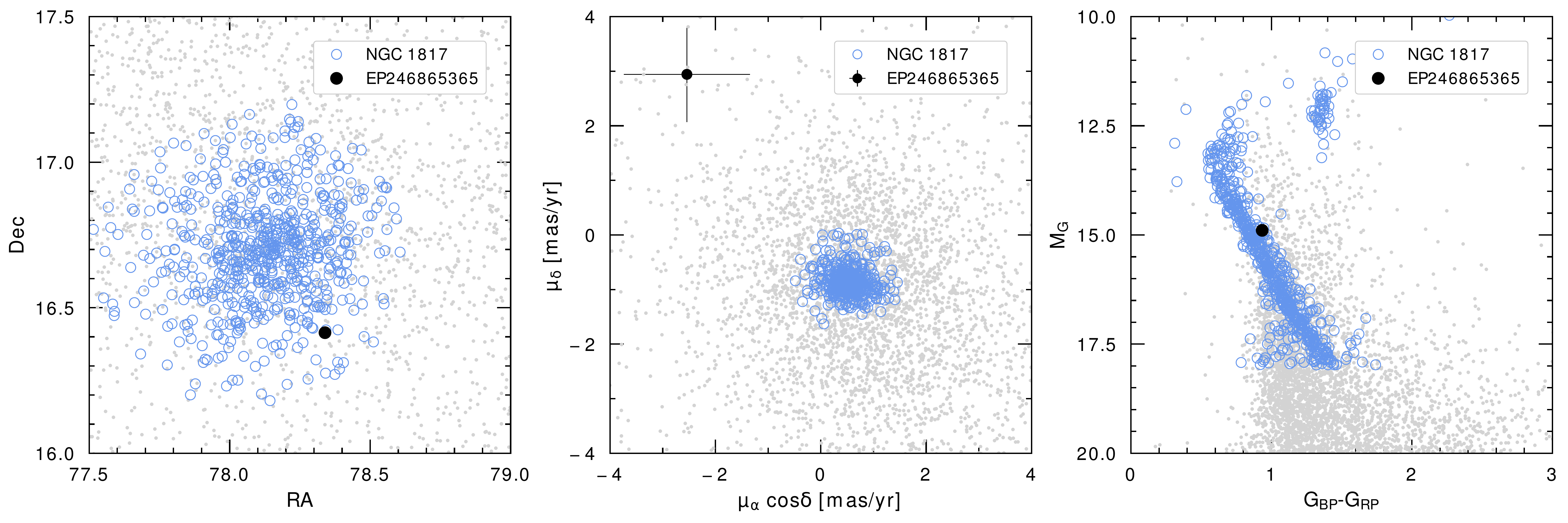}
   \caption{
   \textit{Left}: RA and Dec of stars in the region of interest. NGC 1817 cluster members denoted with blue unfilled circles. EPIC 246465365 denoted with filled black circle.  \textit{Middle}: Proper motions of stars in the region of interest. NGC 1817 cluster members as identified by \cite{2018GaiaCluster} denoted with blue unfilled circles. EPIC 246465365 denoted with filled black circle. \textit{Right}: Color-Magnitude Diagram of NGC 1817 denoted with blue unfilled circles as identified by \cite{2018GaiaCluster} and EPIC 246465365 denoted with filled black circle.}
   \label{pm}
\end{figure*}

However, the reported \Gaia\ DR2 proper motion of ($-$2.54 $\pm$ 1.19, 2.94 $\pm$ 0.88) mas yr\textsuperscript{-1} in (RA, DEC) for this star suggests the star is not associated with the cluster as shown in Figure \ref{pm}. Additionally, our TRES and McDonald spectra indicate an absolute RV of 9 km s\textsuperscript{-1}, strongly indicating that \thisstar\ is unassociated with NGC 1817.

\subsection{False Positive Probability}
It is important to acknowledge that the source of the transit could be due to other astrophysical sources. This is especially true in crowded fields such as this one, where there are multiple stars within the \textit{K2} aperture. This conundrum motivated our follow-up observations which we have described throughout the paper. Here, we summarize the observations and describe the validation tests we use to show that with high likelihood, EPIC 246865365 is indeed a planet-star system. 

From the seeing-limited photometry results, we were able to determine that the signal was indeed coming from the target, eliminating the other two stars in the aperture as potential contaminants (Figure \ref{lc}). Based on our analysis with the TRES and McDonald spectra, the radial velocities rule out the scenario in which the target is an eclipsing binary system (Figure \ref{rv}). Robo-AO results also reveal that there are no neighboring stars $\leq\Delta$2mag close enough to the target to contaminate its signal (Figure \ref{fov}). However, this does not necessarily rule out \textit{all} cases of potential nearby star contamination, so we statistically validate the system as well to affirm that EPIC 246865365 is the source of the transits. We calculate the false positive probability (FPP) of \thisplanet\ using the VESPA software package \citep{morton2015}, an open-source implementation of the method developed by by \cite{morton12}. Given inputs like the transit shape, orbital period, host stellar parameters, and other observational constraints, VESPA calculates the FPP of transiting planet candidates by considering potential false-positive scenarios such as blends with hierarchical (physically associated) eclipsing binary systems, and coincidentally aligned background eclipsing binary systems. After imposing our constraints on the system, VESPA returned an FPP of 0.01\%. We therefore consider the planet candidate statistically validated.

\section{Discussion}\label{discussion}

In this work, we have shown that there is a hot Saturn orbiting EPIC 246865365. This star was long thought to be a member of the open cluster NGC 1817 due to similar proper motions and positions. However, when we conducted follow-up observations, we found that while the planet is likely real, the star is likely not a member of the cluster. Prior to the release of \Gaia\ DR2, we found the star was not a member due to the radial velocities we measured ($\approx$ 9 km s\textsuperscript{-1}) compared to those reported for the cluster (65.3 km s\textsuperscript{-1}). With the release of \Gaia\ DR2, we have better measured kinematics for both the EPIC246865365 and NGC 1817. We found that the star and cluster do not have common radial velocities or proper motions despite what was previously reported, confirming our findings that this star was not a cluster member. 

This work demonstrates the important role \Gaia\ will play in the search for planets in clusters. While we believe that the known cluster planets do indeed orbit cluster members (since follow-up of the planets included RV measurements confirming membership), the full sample of potential planet hosts in clusters still remains to be established. With \Gaia, we will improve cluster membership probabilities allowing us to create a more accurate and robust sample of cluster stars to search. This, in turn, will improve cluster planet occurrence rates. Additionally, these new membership lists would give us the opportunity to determine if previously unidentified cluster members had planets discovered around them serendipitously in earlier planet hunting surveys. 

One of the key elements of our follow-up observations was seeing-limited imaging. When initially performing photometry on the \textit{\textit{K2}} data, there were three stars within the aperture leaving the transit host unknown. Using KeplerCam, we performed seeing-limited photometry to successfully determine the target as the host star and eliminate potential contaminant sources. We expect a similar problem with the Transiting Exoplanet Survey Satellite (\textit{\textit{TESS}}) mission as the cameras' large pixels will also result in multiple stars with in an aperture \citep{ricker}. Seeing limited follow-up is already being used with \textit{TESS} (e.g. \citealt{vanderspek}, \citealt{gunther}), but the challenge will be even greater in cluster regions. For example, about 2-4 stars from NGC 1817 would be found in just one \textit{TESS} pixel.

The fact that \thisstar\ is not a member of NGC 1817 is another null result in the search for transiting giant planets in clusters. Despite the many attempted targeted surveys in the past to detect such planet systems in various open clusters (e.g. \citealt{47tuc}; \citealt{PepperGaudi}; \citealt{BeattyGaudi}; \citealt{AigrainPont}), a successful transit detection of a giant planet in a cluster remains elusive. This has be attributed to the intrinsically low occurrence rate of giant planets and a paucity of stars in open clusters, leading to a low probability of a detection \citep{vansadersgaudi}. \textit{TESS} will be instrumental in continuing this search as it will observe a number of clusters and will be sensitive to such giant planets. With new membership information from \Gaia\ and new data from \textit{TESS}, perhaps the first transiting giant planet in an open cluster will soon be found.


\acknowledgments
We thank the anonymous referee for a quick and constructive review.

RR gratefully acknowledges the support of the Columbia University Bridge to the Ph.D. Program in STEM.
AV's work was performed in part under contract with the California Institute of Technology/Jet Propulsion Laboratory funded by NASA through the Sagan Fellowship Program executed by the NASA Exoplanet Science Institute. C.B. acknowledges support from the Alfred P. Sloan Foundation.
D.W.L. acknowledges partial support from the \textit{TESS} mission through a sub-award from the Massachusetts Institute of Technology to the Smithsonian Astrophysical Observatory (SAO) and from the \Kepler\ mission under NASA Cooperative agreement NNX13AB58A with SAO.

This research has made use of NASA's Astrophysics Data System and the NASA Exoplanet Archive, which is operated by the California Institute of Technology, under contract with the National Aeronautics and Space Administration under the Exoplanet Exploration Program. The National Geographic Society--Palomar Observatory Sky Atlas (POSS-I) was made by the California Institute of Technology with grants from the National Geographic Society. The Oschin Schmidt Telescope is operated by the California Institute of Technology and Palomar Observatory.

This paper includes data collected by the \Kepler\ mission. Funding for the \Kepler\ mission is provided by the NASA Science Mission directorate. Some of the data presented in this paper were obtained from the Mikulski Archive for Space Telescopes (MAST). STScI is operated by the Association of Universities for Research in Astronomy, Inc., under NASA contract NAS5--26555. Support for MAST for non--HST data is provided by the NASA Office of Space Science via grant NNX13AC07G and by other grants and contracts. This work has made use of data from the European Space Agency (ESA) mission {\it Gaia} (\url{https://www.cosmos.esa.int/gaia}), processed by the {\it Gaia} Data Processing and Analysis Consortium (DPAC, \url{https://www.cosmos.esa.int/web/gaia/dpac/consortium}). Funding for the DPAC has been provided by national institutions, in particular the institutions participating in the {\it Gaia} Multilateral Agreement.

The Robo-AO instrument was developed with support from the National Science Foundation under grants AST-0906060, AST-0960343, and AST-1207891, IUCAA, the Mt. Cuba Astronomical Foundation, and by a gift from Samuel Oschin. The Robo-AO team thanks NSF and NOAO for making the Kitt Peak 2.1-m telescope available. We thank the observatory staff at Kitt Peak for their efforts to assist Robo-AO KP operations. Robo-AO KP is a partnership between the California Institute of Technology, the University of Hawai`i, the University of North Carolina at Chapel Hill, the Inter-University Centre for Astronomy and Astrophysics (IUCAA) at Pune, India, and the National Central University, Taiwan. The Murty family feels very happy to have added a small value to this important project. Robo-AO KP is also supported by grants from the John Templeton Foundation and the Mt. Cuba Astronomical Foundation. 

Some data are based on observations at Kitt Peak National Observatory, National Optical Astronomy Observatory (NOAO Prop. ID: 15B-3001), which is operated by the Association of Universities for Research in Astronomy (AURA) under cooperative agreement with the National Science Foundation. We are honored to be permitted to conduct observations on Iolkam Du’ag (Kitt Peak), a mountain within the Tohono O'odham Nation with particular significance to the Tohono O'odham people.




%

\vspace{5mm}
\facilities{\Kepler/\textit{K2}, FLWO:1.5m (TRES), FLWO:1.2m (KeplerCam), KPNO2.1m (Robo-AO), Smith (Tull), \Gaia, Exoplanet Archive, MAST, CDS, ADS}


\software{AstroImageJ \citep{AIJ},
          astropy \citep{astropy:2013},  
          batman \citep{batman}, 
          emcee \citep{emcee},
          VESPA \citep{morton2015}
          }

\begin{deluxetable*}{lcccc}
\tablecaption{System Parameters for \thisstar \label{bigtable}}
\tablewidth{0pt}
\tablehead{
  \colhead{Parameter} & 
  \colhead{Value}     &
  \colhead{} &
  \colhead{68.3\% Confidence}     &
  \colhead{Comment}   \\
  \colhead{} & 
  \colhead{}     &
  \colhead{} &
  \colhead{Interval Width}     &
  \colhead{}  
}
\startdata
\emph{Other Designations} & & & \\
Gaia DR2 3393982701757444352  & & & \\
2MASS J05132138+1624510  & & & \\
WISE J051321.40+162451.1 & & & \\
\\
\emph{Basic Information} & & & \\
Right Ascension & 05:13:21.39 & & & A \\
Declination & +16:24:51.13 & & & A \\
Proper Motion in RA [\ensuremath{\rm mas\,yr^{-1}}]& -2.54 & $\pm$ & 1.19&A  \\
Proper Motion in Dec [\ensuremath{\rm mas\,yr^{-1}}]& 2.94 & $\pm$ & 0.88&A  \\
Absolute Radial Velocity [\kms]& 8.97 & $\pm$ & 1.79 & B \\
$V$-magnitude & 14.96 &$\pm$  & 0.04 & A\\ 
$K$-magnitude & 13.20 &$\pm$  & 0.02 & A\\ 
\Kepler-band Kp magnitude & 14.75 & &  & A\\
\\
\emph{Stellar Parameters} & & & \\
Mass $M_\star$~[$M_\odot$] & \mstar & $\pm$&$ \mstare$ & C,D \\
Radius $R_\star$~[$R_\odot$] & \rstar & $\pm$&$ \rstare$ & C,D \\
Limb darkening $u_1$~ & \ldone  & $\pm$&$ \uldone$ & D,E \\
Limb darkening $u_2$~ & \ldtwo  & $\pm$&$ \uldtwo$ & D,E \\
$\log g_{\rm Kea}$~[cgs] & \loggspc & $\pm$& \loggespc & C \\
Metallicity \metallicity & \mh & $\pm$&\mhe & C \\
$T_{\rm eff}$ [K] & 6100 & $\pm$& 263 & C\\
& & & \\
\emph{\thisplanet} & & & \\
Orbital Period, $P$~[days] & \perplb & $\pm$ &$ \uperplb $ & D \\
Radius Ratio, $R_P/R_\star$ & \rprstb & $\pm$ &$ \urprstb$ & D \\
Scaled semimajor axis, $a/R_\star$  & \arstb & $\pm$ &$ \uarstb$ & D \\
Orbital inclination, $i$~[deg] & \inclb & &$ \uinclb$ & D \\
Transit impact parameter, $b$ & \impb & &$ \uimpb$ & D \\
Transit Duration, $t_{14}$~[hours] & $\tdurb$ & & $\utdurb$ & D \\
Time of Transit $t_{t}$~[BJD] & \ttransitb &$\pm$ & \uttransitb & D\\ 
Planet Radius $R_P$~[\rearth] & \rplb &   $\pm$& $\urplb$  & C,D \\
Planet Radius $R_P$~[$\text{R}_{\text{Jupiter}}$] & \rplbl &   $\pm$& $\urplbl$  & C,D \\
\enddata
\tablecomments{A: Parameters come from the EPIC catalog \citep{epic} and \Gaia\ Data Release 2 \citep{gaiadr2}. B: Parameters come from analysis of the two TRES spectra and three McDonald spectra (Section \ref{rvanalysis}). C: Parameters come from analysis of McDonald spectra using \textit{Kea} (Section \ref{stellarparameters}). D: Parameters come from analysis of the \textit{K2} and KeplerCam light curves (Section \ref{transitanalysis}). E: Gaussian priors of imposed on $u_1$ and $u_2$ centered on 0.324 and 0.299, respectively, with width 0.1 from \cite{claretbloemen}.}
\end{deluxetable*}

\bibliographystyle{aasjournal}

\begin{thebibliography}{}
\expandafter\ifx\csname natexlab\endcsname\relax\def\natexlab#1{#1}\fi
\providecommand{\url}[1]{\href{#1}{#1}}
\providecommand{\dodoi}[1]{doi:~\href{http://doi.org/#1}{\nolinkurl{#1}}}
\providecommand{\doeprint}[1]{\href{http://ascl.net/#1}{\nolinkurl{http://ascl.net/#1}}}
\providecommand{\doarXiv}[1]{\href{https://arxiv.org/abs/#1}{\nolinkurl{https://arxiv.org/abs/#1}}}

\bibitem[{{Aigrain} {et~al.}(2007){Aigrain}, {Hodgkin}, {Irwin}, {Hebb},
  {Irwin}, {Favata}, {Moraux}, \& {Pont}}]{AigrainPont}
{Aigrain}, S., {Hodgkin}, S., {Irwin}, J., {et~al.} 2007, \mnras, 375, 29,
  \dodoi{10.1111/j.1365-2966.2006.11303.x}

\bibitem[{{Astropy Collaboration} {et~al.}(2013){Astropy Collaboration},
  {Robitaille}, {Tollerud}, {Greenfield}, {Droettboom}, {Bray}, {Aldcroft},
  {Davis}, {Ginsburg}, {Price-Whelan}, {Kerzendorf}, {Conley}, {Crighton},
  {Barbary}, {Muna}, {Ferguson}, {Grollier}, {Parikh}, {Nair}, {Unther},
  {Deil}, {Woillez}, {Conseil}, {Kramer}, {Turner}, {Singer}, {Fox}, {Weaver},
  {Zabalza}, {Edwards}, {Azalee Bostroem}, {Burke}, {Casey}, {Crawford},
  {Dencheva}, {Ely}, {Jenness}, {Labrie}, {Lim}, {Pierfederici}, {Pontzen},
  {Ptak}, {Refsdal}, {Servillat}, \& {Streicher}}]{astropy:2013}
{Astropy Collaboration}, {Robitaille}, T.~P., {Tollerud}, E.~J., {et~al.} 2013,
  \aap, 558, A33, \dodoi{10.1051/0004-6361/201322068}

\bibitem[{{Balaguer-N{\'u}{\~n}ez} {et~al.}(2004){Balaguer-N{\'u}{\~n}ez},
  {Jordi}, {Galad{\'\i}-Enr{\'\i}quez}, \& {Zhao}}]{2004balaguer}
{Balaguer-N{\'u}{\~n}ez}, L., {Jordi}, C., {Galad{\'\i}-Enr{\'\i}quez}, D., \&
  {Zhao}, J.~L. 2004, \aap, 426, 819, \dodoi{10.1051/0004-6361:20041332}

\bibitem[{{Baranec} {et~al.}(2014){Baranec}, {Riddle}, {Law}, {Ramaprakash},
  {Tendulkar}, {Hogstrom}, {Bui}, {Burse}, {Chordia}, {Das}, {Dekany},
  {Kulkarni}, \& {Punnadi}}]{Baranec2014}
{Baranec}, C., {Riddle}, R., {Law}, N.~M., {et~al.} 2014, \apjl, 790, L8,
  \dodoi{10.1088/2041-8205/790/1/L8}

\bibitem[{{Beatty} \& {Gaudi}(2008)}]{BeattyGaudi}
{Beatty}, T.~G., \& {Gaudi}, B.~S. 2008, \apj, 686, 1302,
  \dodoi{10.1086/591441}

\bibitem[{{Cantat-Gaudin} {et~al.}(2018){Cantat-Gaudin}, {Jordi}, {Vallenari},
  {Bragaglia}, {Balaguer-N{\'u}{\~n}ez}, {Soubiran}, {Bossini}, {Moitinho},
  {Castro-Ginard}, {Krone-Martins}, {Casamiquela}, {Sordo}, \&
  {Carrera}}]{2018GaiaCluster}
{Cantat-Gaudin}, T., {Jordi}, C., {Vallenari}, A., {et~al.} 2018, \aap, 618,
  A93, \dodoi{10.1051/0004-6361/201833476}

\bibitem[{{Carter} {et~al.}(2011){Carter}, {Winn}, {Holman}, {Fabrycky},
  {Berta}, {Burke}, \& {Nutzman}}]{idlred}
{Carter}, J.~A., {Winn}, J.~N., {Holman}, M.~J., {et~al.} 2011, \apj, 730, 82,
  \dodoi{10.1088/0004-637X/730/2/82}

\bibitem[{{Claret} \& {Bloemen}(2011)}]{claretbloemen}
{Claret}, A., \& {Bloemen}, S. 2011, \aap, 529, A75,
  \dodoi{10.1051/0004-6361/201116451}

\bibitem[{{Collins} {et~al.}(2017){Collins}, {Kielkopf}, {Stassun}, \&
  {Hessman}}]{AIJ}
{Collins}, K.~A., {Kielkopf}, J.~F., {Stassun}, K.~G., \& {Hessman}, F.~V.
  2017, \aj, 153, 77, \dodoi{10.3847/1538-3881/153/2/77}

\bibitem[{{Crossfield} {et~al.}(2015){Crossfield}, {Petigura}, {Schlieder},
  {Howard}, {Fulton}, {Aller}, {Ciardi}, {L{\'e}pine}, {Barclay}, {de Pater},
  {de Kleer}, {Quintana}, {Christiansen}, {Schlafly}, {Kaltenegger}, {Crepp},
  {Henning}, {Obermeier}, {Deacon}, {Weiss}, {Isaacson}, {Hansen}, {Liu},
  {Greene}, {Howell}, {Barman}, \& {Mordasini}}]{Crossfield}
{Crossfield}, I. J.~M., {Petigura}, E., {Schlieder}, J.~E., {et~al.} 2015,
  \apj, 804, 10, \dodoi{10.1088/0004-637X/804/1/10}

\bibitem[{{Cumming} {et~al.}(1999){Cumming}, {Marcy}, \& {Butler}}]{rvcalc}
{Cumming}, A., {Marcy}, G.~W., \& {Butler}, R.~P. 1999, \apj, 526, 890,
  \dodoi{10.1086/308020}

\bibitem[{{Curtis} {et~al.}(2018){Curtis}, {Vanderburg}, {Torres}, {Kraus},
  {Huber}, {Mann}, {Rizzuto}, {Isaacson}, {Howard}, {Henze}, {Fulton}, \&
  {Wright}}]{ruprecht}
{Curtis}, J.~L., {Vanderburg}, A., {Torres}, G., {et~al.} 2018, \aj, 155, 173,
  \dodoi{10.3847/1538-3881/aab49c}

\bibitem[{{Donati} {et~al.}(2014){Donati}, {Beccari}, {Bragaglia}, {Cignoni},
  \& {Tosi}}]{2013Donati}
{Donati}, P., {Beccari}, G., {Bragaglia}, A., {Cignoni}, M., \& {Tosi}, M.
  2014, \mnras, 437, 1241, \dodoi{10.1093/mnras/stt1944}

\bibitem[{{Endl} \& {Cochran}(2016)}]{2016Endl}
{Endl}, M., \& {Cochran}, W.~D. 2016, Publications of the Astronomical Society
  of the Pacific, 128, 094502, \dodoi{10.1088/1538-3873/128/967/094502}

\bibitem[{{Foreman-Mackey} {et~al.}(2013){Foreman-Mackey}, {Hogg}, {Lang}, \&
  {Goodman}}]{emcee}
{Foreman-Mackey}, D., {Hogg}, D.~W., {Lang}, D., \& {Goodman}, J. 2013,
  Publications of the Astronomical Society of the Pacific, 125, 306,
  \dodoi{10.1086/670067}

\bibitem[{{Gaia Collaboration} {et~al.}(2018){Gaia Collaboration}, {Brown},
  {Vallenari}, {Prusti}, {de Bruijne}, {Babusiaux}, {Bailer-Jones}, {Biermann},
  {Evans}, {Eyer}, {Jansen}, {Jordi}, {Klioner}, {Lammers}, {Lindegren},
  {Luri}, {Mignard}, {Panem}, {Pourbaix}, {Randich}, {Sartoretti}, {Siddiqui},
  {Soubiran}, {van Leeuwen}, {Walton}, {Arenou}, {Bastian}, {Cropper},
  {Drimmel}, {Katz}, {Lattanzi}, {Bakker}, {Cacciari}, {Casta{\~n}eda},
  {Chaoul}, {Cheek}, {De Angeli}, {Fabricius}, {Guerra}, {Holl}, {Masana},
  {Messineo}, {Mowlavi}, {Nienartowicz}, {Panuzzo}, {Portell}, {Riello},
  {Seabroke}, {Tanga}, {Th{\'e}venin}, {Gracia-Abril}, {Comoretto},
  {Garcia-Reinaldos}, {Teyssier}, {Altmann}, {Andrae}, {Audard},
  {Bellas-Velidis}, {Benson}, {Berthier}, {Blomme}, {Burgess}, {Busso},
  {Carry}, {Cellino}, {Clementini}, {Clotet}, {Creevey}, {Davidson}, {De
  Ridder}, {Delchambre}, {Dell'Oro}, {Ducourant},
  {Fern{\'a}ndez-Hern{\'a}ndez}, {Fouesneau}, {Fr{\'e}mat}, {Galluccio},
  {Garc{\'\i}a-Torres}, {Gonz{\'a}lez-N{\'u}{\~n}ez}, {Gonz{\'a}lez-Vidal},
  {Gosset}, {Guy}, {Halbwachs}, {Hambly}, {Harrison}, {Hern{\'a}ndez},
  {Hestroffer}, {Hodgkin}, {Hutton}, {Jasniewicz}, {Jean-Antoine-Piccolo},
  {Jordan}, {Korn}, {Krone-Martins}, {Lanzafame}, {Lebzelter}, {L{\"o}ffler},
  {Manteiga}, {Marrese}, {Mart{\'\i}n-Fleitas}, {Moitinho}, {Mora}, {Muinonen},
  {Osinde}, {Pancino}, {Pauwels}, {Petit}, {Recio-Blanco}, {Richards},
  {Rimoldini}, {Robin}, {Sarro}, {Siopis}, {Smith}, {Sozzetti}, {S{\"u}veges},
  {Torra}, {van Reeven}, {Abbas}, {Abreu Aramburu}, {Accart}, {Aerts},
  {Altavilla}, {{\'A}lvarez}, {Alvarez}, {Alves}, {Anderson}, {Andrei},
  {Anglada Varela}, {Antiche}, {Antoja}, {Arcay}, {Astraatmadja}, {Bach},
  {Baker}, {Balaguer-N{\'u}{\~n}ez}, {Balm}, {Barache}, {Barata}, {Barbato},
  {Barblan}, {Barklem}, {Barrado}, {Barros}, {Barstow}, {Bartholom{\'e}
  Mu{\~n}oz}, {Bassilana}, {Becciani}, {Bellazzini}, {Berihuete}, {Bertone},
  {Bianchi}, {Bienaym{\'e}}, {Blanco-Cuaresma}, {Boch}, {Boeche}, {Bombrun},
  {Borrachero}, {Bossini}, {Bouquillon}, {Bourda}, {Bragaglia}, {Bramante},
  {Breddels}, {Bressan}, {Brouillet}, {Br{\"u}semeister}, {Brugaletta},
  {Bucciarelli}, {Burlacu}, {Busonero}, {Butkevich}, {Buzzi}, {Caffau},
  {Cancelliere}, {Cannizzaro}, {Cantat-Gaudin}, {Carballo}, {Carlucci},
  {Carrasco}, {Casamiquela}, {Castellani}, {Castro-Ginard}, {Charlot},
  {Chemin}, {Chiavassa}, {Cocozza}, {Costigan}, {Cowell}, {Crifo}, {Crosta},
  {Crowley}, {Cuypers}, {Dafonte}, {Damerdji}, {Dapergolas}, {David}, {David},
  {de Laverny}, {De Luise}, {De March}, {de Martino}, {de Souza}, {de Torres},
  {Debosscher}, {del Pozo}, {Delbo}, {Delgado}, {Delgado}, {Di Matteo},
  {Diakite}, {Diener}, {Distefano}, {Dolding}, {Drazinos}, {Dur{\'a}n},
  {Edvardsson}, {Enke}, {Eriksson}, {Esquej}, {Eynard Bontemps}, {Fabre},
  {Fabrizio}, {Faigler}, {Falc{\~a}o}, {Farr{\`a}s Casas}, {Federici},
  {Fedorets}, {Fernique}, {Figueras}, {Filippi}, {Findeisen}, {Fonti},
  {Fraile}, {Fraser}, {Fr{\'e}zouls}, {Gai}, {Galleti}, {Garabato},
  {Garc{\'\i}a-Sedano}, {Garofalo}, {Garralda}, {Gavel}, {Gavras}, {Gerssen},
  {Geyer}, {Giacobbe}, {Gilmore}, {Girona}, {Giuffrida}, {Glass}, {Gomes},
  {Granvik}, {Gueguen}, {Guerrier}, {Guiraud}, {Guti{\'e}rrez-S{\'a}nchez},
  {Haigron}, {Hatzidimitriou}, {Hauser}, {Haywood}, {Heiter}, {Helmi}, {Heu},
  {Hilger}, {Hobbs}, {Hofmann}, {Holland}, {Huckle}, {Hypki}, {Icardi},
  {Jan{\ss}en}, {Jevardat de Fombelle}, {Jonker}, {Juh{\'a}sz}, {Julbe},
  {Karampelas}, {Kewley}, {Klar}, {Kochoska}, {Kohley}, {Kolenberg},
  {Kontizas}, {Kontizas}, {Koposov}, {Kordopatis}, {Kostrzewa-Rutkowska},
  {Koubsky}, {Lambert}, {Lanza}, {Lasne}, {Lavigne}, {Le Fustec}, {Le
  Poncin-Lafitte}, {Lebreton}, {Leccia}, {Leclerc}, {Lecoeur-Taibi},
  {Lenhardt}, {Leroux}, {Liao}, {Licata}, {Lindstr{\o}m}, {Lister}, {Livanou},
  {Lobel}, {L{\'o}pez}, {Managau}, {Mann}, {Mantelet}, {Marchal}, {Marchant},
  {Marconi}, {Marinoni}, {Marschalk{\'o}}, {Marshall}, {Martino}, {Marton},
  {Mary}, {Massari}, {Matijevi{\v{c}}}, {Mazeh}, {McMillan}, {Messina},
  {Michalik}, {Millar}, {Molina}, {Molinaro}, {Moln{\'a}r}, {Montegriffo},
  {Mor}, {Morbidelli}, {Morel}, {Morris}, {Mulone}, {Muraveva}, {Musella},
  {Nelemans}, {Nicastro}, {Noval}, {O'Mullane}, {Ord{\'e}novic},
  {Ord{\'o}{\~n}ez-Blanco}, {Osborne}, {Pagani}, {Pagano}, {Pailler},
  {Palacin}, {Palaversa}, {Panahi}, {Pawlak}, {Piersimoni}, {Pineau}, {Plachy},
  {Plum}, {Poggio}, {Poujoulet}, {Pr{\v{s}}a}, {Pulone}, {Racero}, {Ragaini},
  {Rambaux}, {Ramos-Lerate}, {Regibo}, {Reyl{\'e}}, {Riclet}, {Ripepi}, {Riva},
  {Rivard}, {Rixon}, {Roegiers}, {Roelens}, {Romero-G{\'o}mez}, {Rowell},
  {Royer}, {Ruiz-Dern}, {Sadowski}, {Sagrist{\`a} Sell{\'e}s}, {Sahlmann},
  {Salgado}, {Salguero}, {Sanna}, {Santana-Ros}, {Sarasso}, {Savietto},
  {Schultheis}, {Sciacca}, {Segol}, {Segovia}, {S{\'e}gransan}, {Shih},
  {Siltala}, {Silva}, {Smart}, {Smith}, {Solano}, {Solitro}, {Sordo}, {Soria
  Nieto}, {Souchay}, {Spagna}, {Spoto}, {Stampa}, {Steele},
  {Steidelm{\"u}ller}, {Stephenson}, {Stoev}, {Suess}, {Surdej}, {Szabados},
  {Szegedi-Elek}, {Tapiador}, {Taris}, {Tauran}, {Taylor}, {Teixeira},
  {Terrett}, {Teyssand ier}, {Thuillot}, {Titarenko}, {Torra Clotet}, {Turon},
  {Ulla}, {Utrilla}, {Uzzi}, {Vaillant}, {Valentini}, {Valette}, {van Elteren},
  {Van Hemelryck}, {van Leeuwen}, {Vaschetto}, {Vecchiato}, {Veljanoski},
  {Viala}, {Vicente}, {Vogt}, {von Essen}, {Voss}, {Votruba}, {Voutsinas},
  {Walmsley}, {Weiler}, {Wertz}, {Wevers}, {Wyrzykowski}, {Yoldas},
  {{\v{Z}}erjal}, {Ziaeepour}, {Zorec}, {Zschocke}, {Zucker}, {Zurbach}, \&
  {Zwitter}}]{gaiadr2}
{Gaia Collaboration}, {Brown}, A.~G.~A., {Vallenari}, A., {et~al.} 2018, \aap,
  616, A1, \dodoi{10.1051/0004-6361/201833051}

\bibitem[{{Gilliland} {et~al.}(2000){Gilliland}, {Brown}, {Guhathakurta},
  {Sarajedini}, {Milone}, {Albrow}, {Baliber}, {Bruntt}, {Burrows},
  {Charbonneau}, {Choi}, {Cochran}, {Edmonds}, {Frandsen}, {Howell}, {Lin},
  {Marcy}, {Mayor}, {Naef}, {Sigurdsson}, {Stagg}, {Vandenberg}, {Vogt}, \&
  {Williams}}]{47tuc}
{Gilliland}, R.~L., {Brown}, T.~M., {Guhathakurta}, P., {et~al.} 2000, \apj,
  545, L47, \dodoi{10.1086/317334}

\bibitem[{{Goldreich} \& {Soter}(1966)}]{goldreich}
{Goldreich}, P., \& {Soter}, S. 1966, \icarus, 5, 375,
  \dodoi{10.1016/0019-1035(66)90051-0}

\bibitem[{{G{\"u}nther} {et~al.}(2019){G{\"u}nther}, {Pozuelos}, {Dittmann},
  {Dragomir}, {Kane}, {Daylan}, {Feinstein}, {Huang}, {Morton}, {Bonfanti},
  {Bouma}, {Burt}, {Collins}, {Lissauer}, {Matthews}, {Vanderburg}, {Wang},
  {Winters}, {Ricker}, {Vanderspek}, {Latham}, {Seager}, {Winn}, {Jenkins},
  {Armstrong}, {Barakoui}, {Batalha}, {Bean}, {Caldwell}, {Ciardi}, {Collins},
  {Crossfield}, {Fausnaugh}, {Furesz}, {Gan}, {Gillon}, {Guerrero}, {Horne},
  {Howell}, {Ireland}, {Isopi}, {Jehin}, {Kielkopf}, {Lepine}, {Mallia},
  {Matson}, {Myers}, {Palle}, {Quinn}, {Relles}, {Rojas-Ayala}, {Schlieder},
  {Sefako}, {Shporer}, {Su{\'a}rez}, {Tan}, {Ting}, {Twicken}, \&
  {Waite}}]{gunther}
{G{\"u}nther}, M.~N., {Pozuelos}, F.~J., {Dittmann}, J.~A., {et~al.} 2019,
  arXiv e-prints, arXiv:1903.06107.
\newblock \doarXiv{1903.06107}

\bibitem[{{Howell} {et~al.}(2014){Howell}, {Sobeck}, {Haas}, {Still},
  {Barclay}, {Mullally}, {Troeltzsch}, {Aigrain}, {Bryson}, {Caldwell},
  {Chaplin}, {Cochran}, {Huber}, {Marcy}, {Miglio}, {Najita}, {Smith},
  {Twicken}, \& {Fortney}}]{howell}
{Howell}, S.~B., {Sobeck}, C., {Haas}, M., {et~al.} 2014, \pasp, 126, 398,
  \dodoi{10.1086/676406}

\bibitem[{{Huber} {et~al.}(2016){Huber}, {Bryson}, {Haas}, {Barclay},
  {Barentsen}, {Howell}, {Sharma}, {Stello}, \& {Thompson}}]{epic}
{Huber}, D., {Bryson}, S.~T., {Haas}, M.~R., {et~al.} 2016, \apjs, 224, 2,
  \dodoi{10.3847/0067-0049/224/1/2}

\bibitem[{{Jensen-Clem} {et~al.}(2018){Jensen-Clem}, {Duev}, {Riddle},
  {Salama}, {Baranec}, {Law}, {Kulkarni}, \& {Ramprakash}}]{JensenClem2018}
{Jensen-Clem}, R., {Duev}, D.~A., {Riddle}, R., {et~al.} 2018, \aj, 155, 32,
  \dodoi{10.3847/1538-3881/aa9be6}

\bibitem[{{Kipping}(2010)}]{binning}
{Kipping}, D.~M. 2010, \mnras, 408, 1758,
  \dodoi{10.1111/j.1365-2966.2010.17242.x}

\bibitem[{{Kov{\'a}cs} {et~al.}(2002){Kov{\'a}cs}, {Zucker}, \&
  {Mazeh}}]{kovacs}
{Kov{\'a}cs}, G., {Zucker}, S., \& {Mazeh}, T. 2002, \aap, 391, 369,
  \dodoi{10.1051/0004-6361:20020802}

\bibitem[{{Kreidberg}(2015)}]{batman}
{Kreidberg}, L. 2015, Publications of the Astronomical Society of the Pacific,
  127, 1161, \dodoi{10.1086/683602}

\bibitem[{{Krone-Martins} {et~al.}(2010){Krone-Martins}, {Soubiran},
  {Ducourant}, {Teixeira}, \& {Le Campion}}]{2010KroneMartins}
{Krone-Martins}, A., {Soubiran}, C., {Ducourant}, C., {Teixeira}, R., \& {Le
  Campion}, J.~F. 2010, \aap, 516, A3, \dodoi{10.1051/0004-6361/200913881}

\bibitem[{Lammer {et~al.}(2014)Lammer, Stökl, Erkaev, Dorfi, Odert, Güdel,
  Kulikov, Kislyakova, \& Leitzinger}]{Lammer}
Lammer, H., Stökl, A., Erkaev, N.~V., {et~al.} 2014, Monthly Notices of the
  Royal Astronomical Society, 439, 3225, \dodoi{10.1093/mnras/stu085}

\bibitem[{{Law} {et~al.}(2014){Law}, {Morton}, {Baranec}, {Riddle},
  {Ravichandran}, {Ziegler}, {Johnson}, {Tendulkar}, {Bui}, {Burse}, {Das},
  {Dekany}, {Kulkarni}, {Punnadi}, \& {Ramaprakash}}]{Law2014}
{Law}, N.~M., {Morton}, T., {Baranec}, C., {et~al.} 2014, \apj, 791, 35,
  \dodoi{10.1088/0004-637X/791/1/35}

\bibitem[{{Lopez} {et~al.}(2012){Lopez}, {Fortney}, \& {Miller}}]{2012Lopez}
{Lopez}, E.~D., {Fortney}, J.~J., \& {Miller}, N. 2012, \apj, 761, 59,
  \dodoi{10.1088/0004-637X/761/1/59}

\bibitem[{{Mandel} \& {Agol}(2002)}]{mandelagol}
{Mandel}, K., \& {Agol}, E. 2002, \apjl, 580, L171, \dodoi{10.1086/345520}

\bibitem[{{Mann} {et~al.}(2010){Mann}, {Gaidos}, \& {Gaudi}}]{2010Mann}
{Mann}, A.~W., {Gaidos}, E., \& {Gaudi}, B.~S. 2010, \apj, 719, 1454,
  \dodoi{10.1088/0004-637X/719/2/1454}

\bibitem[{{Mann} {et~al.}(2016){Mann}, {Gaidos}, {Mace}, {Johnson}, {Bowler},
  {LaCourse}, {Jacobs}, {Vanderburg}, {Kraus}, {Kaplan}, \& {Jaffe}}]{2016Mann}
{Mann}, A.~W., {Gaidos}, E., {Mace}, G.~N., {et~al.} 2016, \apj, 818, 46,
  \dodoi{10.3847/0004-637X/818/1/46}

\bibitem[{{Meibom} {et~al.}(2013){Meibom}, {Torres}, {Fressin}, {Latham},
  {Rowe}, {Ciardi}, {Bryson}, {Rogers}, {Henze}, {Janes}, {Barnes}, {Marcy},
  {Isaacson}, {Fischer}, {Howell}, {Horch}, {Jenkins}, {Schuler}, \&
  {Crepp}}]{meibom}
{Meibom}, S., {Torres}, G., {Fressin}, F., {et~al.} 2013, \nat, 499, 55,
  \dodoi{10.1038/nature12279}

\bibitem[{{Mills} \& {Fabrycky}(2017)}]{Mills}
{Mills}, S.~M., \& {Fabrycky}, D.~C. 2017, \apj, 838, L11,
  \dodoi{10.3847/2041-8213/aa6543}

\bibitem[{{Morton}(2012)}]{morton12}
{Morton}, T.~D. 2012, \apj, 761, 6, \dodoi{10.1088/0004-637X/761/1/6}

\bibitem[{{Morton}(2015)}]{morton2015}
---. 2015, {VESPA: False positive probabilities calculator}, Astrophysics
  Source Code Library.
\newblock \doeprint{1503.011}

\bibitem[{{Pepper} \& {Gaudi}(2005)}]{PepperGaudi}
{Pepper}, J., \& {Gaudi}, B.~S. 2005, \apj, 631, 581, \dodoi{10.1086/432532}

\bibitem[{{Pepper} {et~al.}(2017){Pepper}, {Gillen}, {Parviainen},
  {Hillenbrand}, {Cody}, {Aigrain}, {Stauffer}, {Vrba}, {David}, {Lillo-Box},
  {Stassun}, {Conroy}, {Pope}, \& {Barrado}}]{2017Pepper}
{Pepper}, J., {Gillen}, E., {Parviainen}, H., {et~al.} 2017, \aj, 153, 177,
  \dodoi{10.3847/1538-3881/aa62ab}

\bibitem[{{Quinn} {et~al.}(2012){Quinn}, {White}, {Latham}, {Buchhave},
  {Cantrell}, {Dahm}, {F{\'{u}}r{\'e}sz}, {Szentgyorgyi}, {Geary}, {Torres},
  {Bieryla}, {Berlind}, {Calkins}, {Esquerdo}, \& {Stefanik}}]{2012Quinn}
{Quinn}, S.~N., {White}, R.~J., {Latham}, D.~W., {et~al.} 2012, \apj, 756, L33,
  \dodoi{10.1088/2041-8205/756/2/L33}

\bibitem[{{Quinn} {et~al.}(2014){Quinn}, {White}, {Latham}, {Buchhave},
  {Torres}, {Stefanik}, {Berlind}, {Bieryla}, {Calkins}, {Esquerdo},
  {F{\'{u}}r{\'e}sz}, {Geary}, \& {Szentgyorgyi}}]{2014Quinn}
---. 2014, \apj, 787, 27, \dodoi{10.1088/0004-637X/787/1/27}

\bibitem[{{Ricker} {et~al.}(2015){Ricker}, {Winn}, {Vanderspek}, {Latham},
  {Bakos}, {Bean}, {Berta-Thompson}, {Brown}, {Buchhave}, {Butler}, {Butler},
  {Chaplin}, {Charbonneau}, {Christensen-Dalsgaard}, {Clampin}, {Deming},
  {Doty}, {De Lee}, {Dressing}, {Dunham}, {Endl}, {Fressin}, {Ge}, {Henning},
  {Holman}, {Howard}, {Ida}, {Jenkins}, {Jernigan}, {Johnson}, {Kaltenegger},
  {Kawai}, {Kjeldsen}, {Laughlin}, {Levine}, {Lin}, {Lissauer}, {MacQueen},
  {Marcy}, {McCullough}, {Morton}, {Narita}, {Paegert}, {Palle}, {Pepe},
  {Pepper}, {Quirrenbach}, {Rinehart}, {Sasselov}, {Sato}, {Seager},
  {Sozzetti}, {Stassun}, {Sullivan}, {Szentgyorgyi}, {Torres}, {Udry}, \&
  {Villasenor}}]{ricker}
{Ricker}, G.~R., {Winn}, J.~N., {Vanderspek}, R., {et~al.} 2015, Journal of
  Astronomical Telescopes, Instruments, and Systems, 1, 014003,
  \dodoi{10.1117/1.JATIS.1.1.014003}

\bibitem[{{Sato} {et~al.}(2007){Sato}, {Izumiura}, {Toyota}, {Kambe}, {Takeda},
  {Masuda}, {Omiya}, {Murata}, {Itoh}, {Ando}, {Yoshida}, {Ikoma}, {Kokubo}, \&
  {Ida}}]{2007Sato}
{Sato}, B., {Izumiura}, H., {Toyota}, E., {et~al.} 2007, \apj, 661, 527,
  \dodoi{10.1086/513503}

\bibitem[{{Seager} \& {Mall{\'e}n-Ornelas}(2003)}]{seagerornelas}
{Seager}, S., \& {Mall{\'e}n-Ornelas}, G. 2003, \apj, 585, 1038,
  \dodoi{10.1086/346105}

\bibitem[{{Tull} {et~al.}(1995){Tull}, {MacQueen}, {Sneden}, \&
  {Lambert}}]{1995Tull}
{Tull}, R.~G., {MacQueen}, P.~J., {Sneden}, C., \& {Lambert}, D.~L. 1995,
  Publications of the Astronomical Society of the Pacific, 107, 251,
  \dodoi{10.1086/133548}

\bibitem[{{van Saders} \& {Gaudi}(2011)}]{vansadersgaudi}
{van Saders}, J.~L., \& {Gaudi}, B.~S. 2011, \apj, 729, 63,
  \dodoi{10.1088/0004-637X/729/1/63}

\bibitem[{{Vanderburg} \& {Johnson}(2014)}]{vj14}
{Vanderburg}, A., \& {Johnson}, J.~A. 2014, \pasp, 126, 948,
  \dodoi{10.1086/678764}

\bibitem[{{Vanderburg} {et~al.}(2016){Vanderburg}, {Latham}, {Buchhave},
  {Bieryla}, {Berlind}, {Calkins}, {Esquerdo}, {Welsh}, \&
  {Johnson}}]{2016vanderburg}
{Vanderburg}, A., {Latham}, D.~W., {Buchhave}, L.~A., {et~al.} 2016, The
  Astrophysical Journal Supplement Series, 222, 14,
  \dodoi{10.3847/0067-0049/222/1/14}

\bibitem[{{Vanderburg} {et~al.}(2017){Vanderburg}, {Becker}, {Buchhave},
  {Mortier}, {Lopez}, {Malavolta}, {Haywood}, {Latham}, {Charbonneau},
  {L{\'o}pez-Morales}, {Adams}, {Bonomo}, {Bouchy}, {Collier Cameron},
  {Cosentino}, {Di Fabrizio}, {Dumusque}, {Fiorenzano}, {Harutyunyan},
  {Johnson}, {Lorenzi}, {Lovis}, {Mayor}, {Micela}, {Molinari}, {Pedani},
  {Pepe}, {Piotto}, {Phillips}, {Rice}, {Sasselov}, {S{\'e}gransan},
  {Sozzetti}, {Udry}, \& {Watson}}]{determinemasses}
{Vanderburg}, A., {Becker}, J.~C., {Buchhave}, L.~A., {et~al.} 2017, \aj, 154,
  237, \dodoi{10.3847/1538-3881/aa918b}

\bibitem[{{Vanderburg} {et~al.}(2018){Vanderburg}, {Mann}, {Rizzuto},
  {Bieryla}, {Kraus}, {Berlind}, {Calkins}, {Curtis}, {Douglas}, {Esquerdo},
  {Everett}, {Horch}, {Howell}, {Latham}, {Mayo}, {Quinn}, {Scott}, \&
  {Stefanik}}]{2018Vanderburg}
{Vanderburg}, A., {Mann}, A.~W., {Rizzuto}, A., {et~al.} 2018, \aj, 156, 46,
  \dodoi{10.3847/1538-3881/aac894}

\bibitem[{{Vanderspek} {et~al.}(2019){Vanderspek}, {Huang}, {Vanderburg},
  {Ricker}, {Latham}, {Seager}, {Winn}, {Jenkins}, {Burt}, {Dittmann},
  {Newton}, {Quinn}, {Shporer}, {Charbonneau}, {Irwin}, {Ment}, {Winters},
  {Collins}, {Evans}, {Gan}, {Hart}, {Jensen}, {Kielkopf}, {Mao}, {Waalkes},
  {Bouchy}, {Marmier}, {Nielsen}, {Ottoni}, {Pepe}, {S{\'e}gransan}, {Udry},
  {Henry}, {Paredes}, {James}, {Hinojosa}, {Silverstein}, {Palle},
  {Berta-Thompson}, {Crossfield}, {Davies}, {Dragomir}, {Fausnaugh}, {Glidden},
  {Pepper}, {Morgan}, {Rose}, {Twicken}, {Villase{\~n}or}, {Yu}, {Bakos},
  {Bean}, {Buchhave}, {Christensen-Dalsgaard}, {Christiansen}, {Ciardi},
  {Clampin}, {De Lee}, {Deming}, {Doty}, {Jernigan}, {Kaltenegger}, {Lissauer},
  {McCullough}, {Narita}, {Paegert}, {Pal}, {Rinehart}, {Sasselov}, {Sato},
  {Sozzetti}, {Stassun}, \& {Torres}}]{vanderspek}
{Vanderspek}, R., {Huang}, C.~X., {Vanderburg}, A., {et~al.} 2019, \apj, 871,
  L24, \dodoi{10.3847/2041-8213/aafb7a}

\bibitem[{{Wu} {et~al.}(2009){Wu}, {Zhou}, {Ma}, \& {Du}}]{2009Wu}
{Wu}, Z.-Y., {Zhou}, X., {Ma}, J., \& {Du}, C.-H. 2009, \mnras, 399, 2146,
  \dodoi{10.1111/j.1365-2966.2009.15416.x}

\bibitem[{{Yi} {et~al.}(2001){Yi}, {Demarque}, {Kim}, {Lee}, {Ree}, {Lejeune},
  \& {Barnes}}]{yi}
{Yi}, S., {Demarque}, P., {Kim}, Y.-C., {et~al.} 2001, The Astrophysical
  Journal Supplement Series, 136, 417, \dodoi{10.1086/321795}

\end{thebibliography}




\end{document}